\documentclass[%
 aip,
 amsmath,amssymb,
 reprint,%
]{revtex4-1}

\usepackage{graphicx}
\usepackage{dcolumn}
\usepackage{bm}

\usepackage[utf8]{inputenc}
\usepackage[T1]{fontenc}
\usepackage{mathptmx}
\usepackage{etoolbox}

\makeatletter
\def\@email#1#2{%
 \endgroup
 \patchcmd{\titleblock@produce}
  {\frontmatter@RRAPformat}
  {\frontmatter@RRAPformat{\produce@RRAP{*#1\href{mailto:#2}{#2}}}\frontmatter@RRAPformat}
  {}{}
}%
\makeatother
\begin{document}

\preprint{AIP/123-QED}

\title{Widely tunable cavity-enhanced backward difference-frequency generation}
\author{Ming-Yuan Gao}
\author{Yue-Wei Song}
\affiliation{ CAS Key Laboratory of Quantum Information, University of Science and Technology of China, Hefei, Anhui 230026, China}
\affiliation{ CAS Center for Excellence in Quantum Information and Quantum Physics, University of Science and Technology of China, Hefei 230026, China}

\author{Ren-Hui Chen}
\affiliation{ CAS Key Laboratory of Quantum Information, University of Science and Technology of China, Hefei, Anhui 230026, China}
\affiliation{ CAS Center for Excellence in Quantum Information and Quantum Physics, University of Science and Technology of China, Hefei 230026, China}
\affiliation{ Hefei National Laboratory, University of Science and Technology of China, Hefei 230088, China}

\author{Yin-Hai Li}
\affiliation{ CAS Key Laboratory of Quantum Information, University of Science and Technology of China, Hefei, Anhui 230026, China}
\affiliation{ CAS Center for Excellence in Quantum Information and Quantum Physics, University of Science and Technology of China, Hefei 230026, China}

\author{Zhi-Yuan Zhou}
\affiliation{ CAS Key Laboratory of Quantum Information, University of Science and Technology of China, Hefei, Anhui 230026, China}
\affiliation{ CAS Center for Excellence in Quantum Information and Quantum Physics, University of Science and Technology of China, Hefei 230026, China}
\affiliation{ Hefei National Laboratory, University of Science and Technology of China, Hefei 230088, China}
\affiliation{ Corresponding author: zyzhouphy@ustc.edu.cn}

\author{Bao-Sen Shi}
\affiliation{ CAS Key Laboratory of Quantum Information, University of Science and Technology of China, Hefei, Anhui 230026, China}
\affiliation{ CAS Center for Excellence in Quantum Information and Quantum Physics, University of Science and Technology of China, Hefei 230026, China}
\affiliation{ Hefei National Laboratory, University of Science and Technology of China, Hefei 230088, China}
\affiliation{ Corresponding author: drshi@ustc.edu.cn}

\date{\today}

\begin{abstract}
Difference-frequency generation (DFG) is a powerful technique for generating widely tunable infrared radiation. However, conventional phase-matching schemes may require tuning multiple parameters—such as the wavelengths, crystal temperature, crystal angle, and poling period—to achieve wide tunability, which increases the complexity of practical operation. In this work, we employ a backward quasi-phase-matching scheme with distinctive tuning characteristics and demonstrate pump-enhanced continuous-wave DFG output tunable from 1751 nm to 2451 nm (700 nm range) in a bulk crystal. The tuning is achieved solely by varying the pump wavelength and the signal wavelength (less than 5 nm), enabling continuous, rapid, and room-temperature operation. The tuning characteristics, power-scaling behavior, and output stability are experimentally verified with the idler wavelength set at 2000 nm. The approach offers a new paradigm for widely tunable infrared radiation generation and holds promise for applications in spectroscopy and biomedical sensing.
\end{abstract}

\maketitle

Continuous-wave (CW) radiation in the 1.7–5.0 µm spectral region holds significant promise for applications in spectroscopy \cite{1}, gas detection \cite{2}, and biomedicine \cite{3}. The development of high-performance CW sources, characterized by compactness, high efficiency, and wide tunability, will strongly advance these fields. Among these performance metrics, wide tunability is of particular importance, as it offers practical advantages such as enabling reliable detection of multi-component gas mixtures. CW sources based on difference-frequency generation (DFG) are promising candidates to meet these performance demands, with their wide tunability as a key competitive advantage.

Tuning property of a CW DFG system is governed by several factors, including the choice of nonlinear crystal, phase-matching scheme, and tuning method. For birefringent phase matching schemes, tuning is typically achieved by adjusting the pump and/or signal wavelengths, the crystal temperature, and/or its angle. For instance, Pine demonstrated 2.2–4.2 µm DFG output in LiNbO\textsubscript{3} by varying both wavelength and temperature \cite{1}. Seiter et al. achieved 3.16–3.67 µm output in LiNbO\textsubscript{3} by tuning wavelength and crystal angle \cite{4}, while Simon et al. used AgGaS\textsubscript{2} with wavelength and angular tuning to realize 3.155–3.423 µm output \cite{5}.
Quasi-phase matching (QPM) offers some notable advantages, such as access to the largest nonlinear coefficient and the elimination of beam walk-off. In QPM-based DFG, the grating period becomes an additional tuning parameter. Goldberg et al. realized 3.0–4.1 µm DFG in periodically poled LiNbO\textsubscript{3} (PPLN) by rotating the bulk crystal to alter the effective grating period in conjunction with wavelength tuning \cite{6}. Cousin et al. reported 3.16–3.41 µm output using PPLN by tuning wavelength and temperature \cite{2}. Zhao et al. demonstrated a tuning range of 3117.2–3598.8 nm by simultaneously tuning the signal and pump wavelengths, with the PPLN crystal maintained at 160 °C \cite{7}. Høgstedt et al. achieved 2.9–3.4 µm DFG by simultaneously tuning wavelength, grating period, and temperature \cite{8}.

Achieving wide tunability in the above CW DFG typically requires adjusting at least one of the pump or signal wavelengths over a wide range to satisfy energy conservation. Phase matching further necessitates tuning one or more parameters: the other input wavelength, crystal temperature, angle, and grating period. Among these parameters, adjusting multiple poling channels and angle of the crystal can both introduce alignment challenges during practical implementation \cite{4,5,6,8}. Temperature tuning is often slow and may require operation far from room temperature \cite{1,2,8}. The ideal DFG tuning approach would allow rapid and continuous tuning, maintain room temperature operation, minimize alignment complexity, and achieve the widest possible tuning range with the fewest number of adjustable parameters.

Backward difference-frequency generation (BDFG) based on backward QPM (BQPM) \cite{9} presents a promising approach to meet the aforementioned requirements. In backward configuration, one of the interacting waves—either the signal or the idler—propagates in the direction opposite to that of the pump and the remaining wave \cite{10}. This phase-matching scheme has already demonstrated distinct advantages over conventional forward phase matching schemes, where all three waves propagate in the same direction, in applications such as mirrorless backward optical parametric oscillators (OPOs) \cite{11,12} and the generation of quantum light sources \cite{13,14}.
BQPM also exhibits unique tuning characteristics that differ from those of forward phase matching. Specifically, the backward-propagating signal wavelength changes slowly in response to tuning of the pump, while the idler wavelength varies much more rapidly \cite{11}. Consequently, BDFG enables wide idler tunability by combining a widely tunable pump source with a signal source requiring only narrow-range tuning, while simultaneously satisfying both phase-matching and energy-conservation conditions.
As previously noted, achieving widely tunable DFG output always requires wide tuning of either the pump or the signal wavelength. When wavelength tuning is achieved solely by adjusting the pump and signal wavelengths, BDFG significantly relaxes the required tuning range of the signal laser or amplifier—a factor that can limit the practicality of forward DFG schemes, as will be further discussed in the theory model section. Moreover, this configuration enables fast, continuous tuning at room temperature and minimize alignment-related challenges.
Moreover, implementing BQPM requires a short poling period; otherwise, higher-order QPM must be employed \cite{9}. To date, however, no scheme employing a cavity to enhance the efficiency of the BDFG process has been reported.

In this work, we demonstrate the realization of widely tunable cavity-enhanced BDFG. Using a bulk periodically poled potassium titanyl phosphate (PPKTP) crystal with a single poling period and operating at room temperature, we achieve widely tunable DFG output from 1751 nm to 2451 nm (700 nm range) by tuning only the pump and signal wavelengths, without requiring adjustments to temperature, angle, or grating period. The required signal tuning range is less than 5 nm. Cavity enhancement is applied to the pump wave, enabling efficient enhancement over an approximately 126 nm pump tuning range. We characterize the wavelength and temperature tuning behavior at a representative idler wavelength of 2000 nm, as well as the dependence of idler power on pump and signal powers, and its power stability. This experiment demonstrates a new paradigm for generating widely tunable infrared radiation (IR).

Under BQPM, the coupled-amplitude equation for the idler is given by \cite{15,16,17,s1}:
\begin{equation}
\begin{array}{l}
\frac{{d{A_i}}}{{dz}} = \frac{{2i{d_{eff}}\omega _i^2}}{{{k_i}{c^2}}}{A_p}A_s^*{e^{i\Delta kz}},
\end{array}\label{eq1}
\end{equation}
where ${A_q}$ is the amplitude of the wave $q$, with $q = i,{\rm{ }}s,{\rm{ }}p$ corresponding to the forward idler, backward signal, and forward pump waves, respectively; here, the idler refers to the difference-frequency wave, and the pump corresponds to the highest-frequency wave; ${\omega _q}$ and ${k_q}$ are respectively the angular frequency, and wave vector of the wave $q$; ${d_{eff}}$ is the effective nonlinear coefficient \cite{18}; $c$ is the light velocity in vacuum; $\Delta k$ is given by
\begin{equation}
\begin{array}{l}
\Delta k = {k_p} - {k_i} + {k_s} - {k_m},
\end{array}\label{eq2}
\end{equation}
${k_m} = 2\pi m/\Lambda $ is the QPM grating vector, $m$ is the order of the interaction and $\Lambda$ is the poling period of the nonlinear crystal.
Assuming constant pump and signal amplitudes and under the phase-matching condition $\Delta k = 0$, the intensity of the idler wave is obtained by integrating \eqref{eq1} in conjunction with ${I_q} = 2{n_q}{\varepsilon _0}c{\left| {{A_q}} \right|^2}$:
\begin{equation}
\begin{array}{l}
{I_i} = \frac{{8{\pi ^2}d_{eff}^2{L^2}}}{{{n_i}{n_s}{n_p}\lambda _i^2c{\varepsilon _0}}}{I_s}{I_p},
\end{array}\label{eq3}
\end{equation}
where ${I_q}$, ${\lambda _q}$, ${n_j}$ denote the intensity, wavelength, and refractive index of the wave $q$, respectively; ${\varepsilon _0}$ is the vacuum dielectric constant; and $L$ is the length of the nonlinear crystal.
The tuning behavior can be further derived from \eqref{eq2} and ${\omega _p} = {\omega _i} + {\omega _s}$ \cite{11,12}:

\begin{equation}
\begin{array}{l}
\frac{{\partial {\omega _i}}}{{\partial {\omega _p}}} = \frac{{\nu _p^{ - 1} + \nu _s^{ - 1}}}{{\nu _i^{ - 1} + \nu _s^{ - 1}}} = 1 + \varepsilon , \\
\frac{{\partial {\omega _s}}}{{\partial {\omega _p}}} = \frac{{\nu _i^{ - 1} - \nu _p^{ - 1}}}{{\nu _i^{ - 1} + \nu _s^{ - 1}}} =  - \varepsilon , 
\end{array}\label{eq4}
\end{equation}
where ${\nu _q} = {{\partial {\omega _q}} \mathord{\left/
 {\vphantom {{\partial {\omega _q}} {\partial {k_q}}}} \right.
 \kern-\nulldelimiterspace} {\partial {k_q}}}$. For the idler wavelength of 2000 nm and the pump wavelength of 873.8 nm used in this work, the value of $\varepsilon $ is approximately 0.01, based on refractive index data from the Ref.~\onlinecite{19}. From the perspective of phase matching in BDFG, this implies that only a small tuning range of the signal wavelength is needed to generate a broad idler tuning range when the pump wavelength is varied.
Table \ref{tab:1} presents a theoretical comparison of the required tuning ranges of the pump and signal wavelengths for generating DFG idler output from 1751 nm to 2451 nm using BQPM and forward QPM (FQPM) in PPKTP. The comparison assumes wavelength tuning of the pump and signal as the sole degrees of freedom. In the BQPM and FQPM cases, the poling periods are 2.95 µm and 27.90 µm, respectively, corresponding to the 7th and 1st QPM orders. The temperature is set at 25 °C, and both configurations employ Type-0 QPM. Only phase matching ($\Delta k  \approx 0$) is considered in the calculation.
As shown in Table \ref{tab:1}, to generate the same 700 nm idler tuning range, the required pump tuning range for BQPM is approximately one order of magnitude larger than that for FQPM, while the required signal tuning range is nearly two orders of magnitude smaller. Specifically, FQPM requires a signal tuning range exceeding 300 nm, which poses challenges for many commercial laser sources and amplifiers in this wavelength region. In contrast, BQPM achieves the same idler coverage with a signal tuning range of less than 5 nm, a specification readily met by a wide range of commercially available lasers and amplifiers.

\begin{table}[htbp]
\caption{\label{tab:1}Required Wavelength Tuning Ranges for Generating 1751 – 2451 nm Idler }
\begin{ruledtabular}
\begin{tabular}{lcr}
Configuration& Pump ranges (nm)& Signal ranges (nm)\\
\hline
BQPM & 823.328 - 949.379 & 1549.61 - 1554.05 \\
FQPM & 815.005 - 823.402 & 1221.00 - 1554.25 \\
\end{tabular}
\end{ruledtabular}
\end{table}

\begin{figure*}[ht]
\centering\includegraphics[width=17cm]{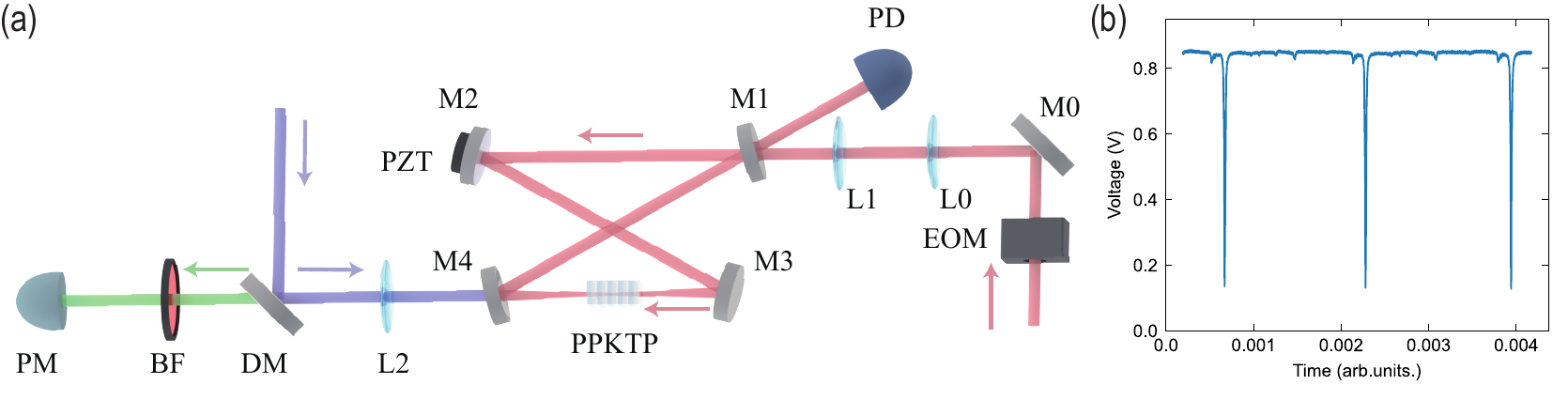}
\caption{(a) Experimental setup. EOM: electro-optic modulator; M0: mirror; M1-M4: cavity mirrors; L: lens; PD: photodetector; PZT: piezoelectric transducer; PPKTP: periodically poled KTP crystal; DM: dichroic mirror; BF: bandpass filter; PM:power meter. (b) Reflection spectrum of the cavity at a pump wavelength of 873.8 nm, recorded on an oscilloscope during cavity-length scanning.}
\label{fig1}
\end{figure*}

The experimental setup is shown in Fig.\ref{fig1}(a). The pump beam is provided by a CW Ti: sapphire laser, which serves as a widely tunable source. Its power is enhanced using a bow-tie ring cavity that is resonant only for the pump. The pump beam is phase-modulated by an EOM to enable Pound–Drever–Hall locking of the cavity. Lenses L0 and L1 are used to mode-match the pump to the cavity. The input coupler M1 has a power reflectivity of approximately 97.5\%, while mirrors M2–M4 are coated for high reflectivity at the pump wavelength. Mirrors M3 and M4 are concave with a radius of curvature of 80 mm. The cavity formed by M1–M4 has a total length of approximately 410 mm. The cavity reflection spectrum is measured by scanning the cavity length using a PZT attached to M2. The reflected signal is detected by a PD and recorded with an oscilloscope. Fig.\ref{fig1}(b) shows the reflection spectrum at a pump wavelength of 873.8 nm, corresponding to the cavity finesse of approximately 162.
The signal beam is generated by a CW diode laser and amplified by an erbium-doped fiber amplifier before being coupled into the cavity. The signal requires only a narrow tuning range. Lens L2 focuses the signal beam into a PPKTP crystal, which is mounted in a temperature-controlled oven. The crystal dimensions are $1 \times 2 \times 10$ $\rm m{m^3}$, with a poling period of 2.95 µm, satisfying 7th-order type-0 BQPM. The generated idler exits from the cavity and is separated using a DM and BF, then its power is measured using a PM.

\begin{figure}[htb]
\centering\includegraphics[width=8.5cm]{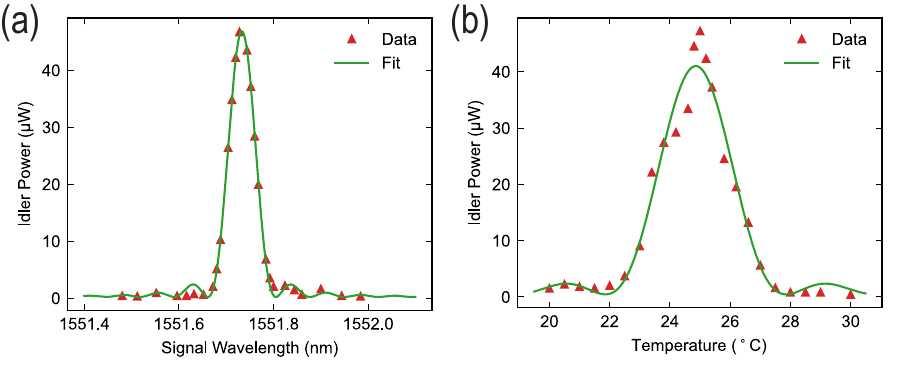}
\caption{(a) Wavelength tuning and (b) temperature tuning measurements of the DFG output. The data are fitted with ${sin}{c^2}$ functions.}
\label{fig2}
\end{figure}

Prior to investigating the wide tunability of the idler, we experimentally characterize its wavelength and temperature tuning behavior, its power dependence on the pump and signal powers, and its output stability at a idler wavelength of 2000 nm.
For the wavelength and temperature tuning measurement, the pump wavelength is fixed at 873.8 nm with an incident power of 200 mW, and the signal power is set to 2.3 W. Wavelength tuning was first performed with the crystal temperature maintained at 25 °C. The signal wavelength is tuned from 1551.48 nm to 1551.98 nm, and the idler power is recorded. The wavelength tuning curve is shown in Fig.\ref{fig2}(a). A ${sin}{c^2}$ function fit yields a bandwidth of 7.9 GHz. A theoretical prediction based on the parameters used in the experiment gives a bandwidth of 7.2 GHz, in good agreement with the measured result.
Temperature tuning was then checked with the signal wavelength fixed at 1551.73 nm. The crystal temperature was varied from 20 °C to 30 °C, and the resulting tuning curve is shown in Fig.\ref{fig2}(b). The measured bandwidth from the ${sin}{c^2}$ fit is 2.7 °C, and the theoretical temperature bandwidth is 1.9 °C. The theoretical calculation method for the wavelength and temperature bandwidths follows Ref.~\onlinecite{9}.

\begin{figure}[htb]
\centering\includegraphics[width=8.5cm]{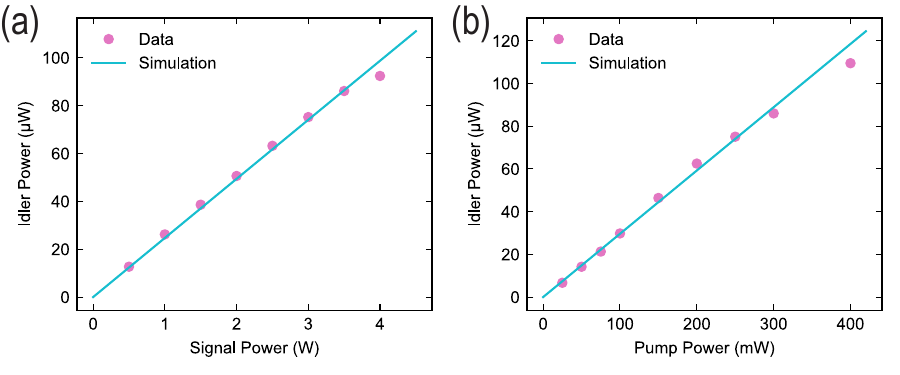}
\caption{Idler power as a function of (a) signal power and (b) pump power. Solid lines indicate simulation results. In (a), the pump power was fixed at 250 mW. In (b), the signal power was fixed at 3 W.}
\label{fig3}
\end{figure}

To evaluate the idler power dependence on the pump and signal powers, the pump wavelength is set to 873.8 nm, the signal wavelength to 1551.73 nm, and the crystal temperature to 25 °C. First, the pump power is fixed at 250 mW, and the signal power is varied from 0.5 W to 4.0 W. The corresponding idler powers are shown in Fig.\ref{fig3}(a). A maximum idler power of 92.4 mW is obtained at 4.0 W signal input.
Next, the signal power is fixed at 3 W while the pump power is varied from 25 mW to 400 mW. The resulting idler output is shown in Fig.\ref{fig3}(b). At a pump power of 400 mW, the idler output reaches 109.5 mW, which corresponds to a power conversion efficiency of 0.009\%, defined as $\eta  = {P_i}/({P_s}{P_p}) \times 100\% $, and a quantum conversion efficiency of 0.02\%, defined as ${\eta _q} = \eta {\lambda _i}/{\lambda _p}$, where ${P_q}$ denotes the power of wave $q$. Optical losses from components affecting the idler were not considered in this calculation.
To obtain the theoretical idler power using \eqref{eq3}, the following assumptions and estimates are made. The cross-sectional areas $A$ of the three interacting beams are assumed to be equal and satisfy \cite{15,20} ${P_q} = {I_q}A$ and $A = \pi {w^2}$, where $w$ is the beam waist radius, estimated to be 52 µm. The mode-matching factor for the current system is estimated to be 0.92, based on the method described in Ref.~\onlinecite{21}. The total transmission loss from the crystal to the PM for the idler is estimated to be approximately 40\%, while the signal power experiences approximately 10\% loss before reaching the crystal. Pump loss is not considered. Using the relation \cite{18} ${d_{eff}} = 2{d_{33}}/7\pi $ with ${d_{33}} = 5.0$ pm/V, \eqref{eq3} in conjunction with cavity-enhancement theory \cite{21,22} yields the simulation curve shown in Fig.\ref{fig3}. The relatively low ${d_{33}}$ could result from imperfections in the fabricated poling period, which can be especially detrimental under higher-order QPM due to increased sensitivity \cite{23,24}. This suggests that the reduction in ${d_{eff}}$ for higher-order QPM compared to first-order QPM not only arises from the intrinsic dependence on the QPM order $m$, but also reflects the greater sensitivity of higher-order processes to imperfections in the poling structure. At high pump powers, deviations between experimental results and simulations may occur, possibly due to increased absorption, thermal effects, or other nonlinear processes \cite{21}.

\begin{figure}[htb]
\centering\includegraphics[width=8.5cm]{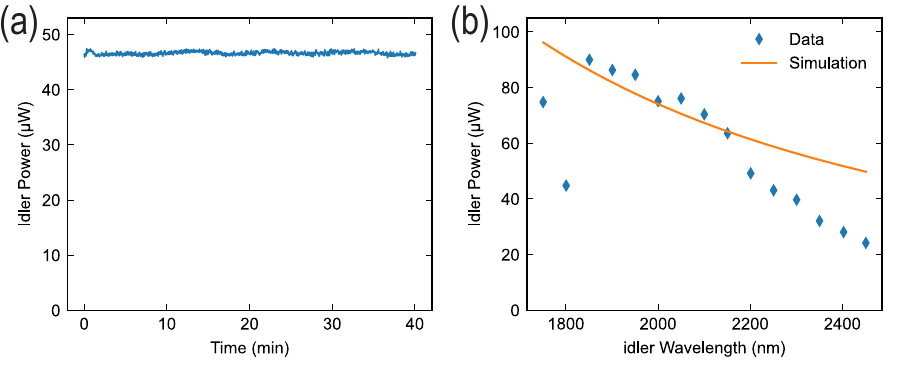}
\caption{(a) Idler power at 2000 nm measured over 40 minutes. The peak-to-peak power stability is 3.6\%. (b) Idler output across the 1751–2451 nm range. The pump and signal powers are fixed at 250 mW and 3 W, respectively. The solid line represents the simulation result.}
\label{fig4}
\end{figure}

The idler power stability is shown in Fig.\ref{fig4}(a), The pump and signal wavelengths are fixed at 873.8 nm and 1551.73 nm, respectively, with powers of 200 mW and 2.3 W. The crystal temperature is maintained at 25 °C. The idler power is recorded over 40 minutes with one data point collected per second. The peak-to-peak power stability, calculated as \begin{math}({P_{{max}}} - {P_{{min}}})/{P_{mean}}\end{math} with \begin{math}{P_{{max}}}\end{math}, \begin{math}{P_{{min}}}\end{math}, and \begin{math}{P_{mean}}\end{math} denoting the maximum, minimum, and average power respectively, is measured to be 3.6\%.

Finally, the wide tunability of the idler output is demonstrated. The pump and signal powers before the cavity are set to 250 mW and 3 W, respectively, and the crystal temperature is maintained at 25 °C. By simultaneously tuning only the pump and signal wavelengths to satisfy both energy conservation and phase matching, idler generation is achieved across a wide spectral range. In the experiment, the pump wavelength is tuned from 823.2 nm to 949.4 nm, covering a range of 126.2 nm, while the signal wavelength is tuned from 1554.03 nm to 1549.59 nm, covering only 4.44 nm. As a result, the generated idler spans from 1751 nm to 2451 nm, yielding a total tuning range of 700 nm. The required tuning ranges of the pump and signal wavelengths agree well with the theoretical values presented in Table \ref{tab:1}. The measured tuning curve, sampled approximately every 50 nm of idler wavelength, is presented in Fig.\ref{fig4}(b).
For the theoretical simulation of the idler power spanning 1751–2451 nm, the following simplifying assumptions are applied: ${{\partial {\lambda _p}} \mathord{\left/
 {\vphantom {{\partial {\lambda _p}} {\partial {\lambda _i}}}} \right.
 \kern-\nulldelimiterspace} {\partial {\lambda _i}}}$ is taken as constant. The beam waist, pump mode-matching factor, and loss estimates for all three interacting waves are assumed to be the same as those used  in the 2000 nm idler case. With ${d_{33}} = 5.0$ pm/V, the simulation curve in Fig.\ref{fig4}(b) is obtained. The significant deviations observed near the spectral edges are likely caused by reduced cavity mode-matching factor and increased optical losses for the idler wave at those wavelengths. In addition, since the pump laser must be tuned over a range of approximately 126 nm to generate the difference-frequency output, it is preferable to use an achromatic wave plate for polarization control; otherwise, degradation of the cavity spectrum may occur.

In summary, we theoretically and experimentally demonstrate cavity-enhanced backward quasi-phase matching (7th-order) enabling continuous, rapid, and widely tunable infrared generation from 1751 nm to 2451 nm at room temperature, using only pump and signal wavelength tuning. At the idler wavelength of 2000 nm, we measure the wavelength and temperature tuning bandwidths to be 7.9 GHz and 2.7 °C, respectively. We also investigate the power dependence of the idler on the pump and signal beams, yielding a maximum conversion efficiency of 0.009\% and a corresponding quantum efficiency of 0.02\%, without accounting for optical losses. The idler output shows stable power over 40 minutes with the peak-to-peak power stability of 3.6\%.
It is worth noting that although mirrorless backward OPOs can also generate similar IR, they are limited to pulsed operation due to prohibitively high thresholds in the CW regime \cite{11}. In contrast, BDFG in the CW domain bypasses this threshold constraint with the simple addition of a narrowband tunable laser, thereby enabling the unique tuning characteristics of BQPM to be exploited in the CW regime. Furthermore, the present system can serve as an extension module for the pump laser (such as a Ti: sapphire laser), offering approximately 1000 nm of infrared tuning range by adding only a single wavelength-tunable laser with a narrow tuning range (approximately 5 nm). This work introduces a promising new approach for IR generation, with potential impact on spectroscopy and biomedical sensing.

\vspace{0.5cm}
 We would like to acknowledge the support from the National Key Research and Development Program of China (2022YFB3903102, 2022YFB3607700), National Natural Science Foundation of China (NSFC)(62435018), Innovation Program for Quantum Science and Technology (2021ZD0301100), USTC Research Funds of the Double First-Class Initiative(YD2030002023), and Research Cooperation Fund of SAST, CASC (SAST2022-075).
\section*{Data Availability Statement}

The data that support the findings of this study are available from the corresponding authors upon reasonable request.

\bibliography{aipsamp}

\end{document}